\documentclass{article}
\usepackage{graphicx}
\usepackage{dcolumn}
\usepackage{amsmath}
\usepackage{bm}
\begin{document}
\title{\bf Particle creation in the framework of $f(G)$ gravity}
\author{R. Rashidi$^{1}$\thanks{email:
reza.rashidi@sru.ac.ir}, F. Ahmadi$^{1}$\thanks{email:
fahmadi@sru.ac.ir} and M. R. Setare$^{2}$\thanks{email:
rezakord@ipm.ir}\\$^1${\small Department of Physics, Shahid
Rajaee Teacher Training University, Lavizan, Tehran 16788, Iran.}\\
$^2${\small Department of Science, Payame Noor University, Bijar, Iran.}\\}
\maketitle

\begin{abstract}
In this paper, we study the problem of massless particle creation in a flat, homogeneous and
isotropic universe in the framework of $f(G)$ gravity. The Bogolyubov coefficients are
calculated for the accelerating power-law solutions of the model
in a matter dominated universe, from which the total number of created
particle per unit volume of space can be obtained. It is proved that the total particle density always has a finite value. Therefore, the Bogolyubov transformations are well-defined and the Hilbert spaces spanned by the vacuum states at different times are unitarily equivalent. We find that the particles with small values of the mode $k$ are produced in the past and particles with large values
of $k$ are produced only in the future. The negative pressure resulting from the gravitational particle creation is also determined. It is then argued that this pressure even in the presence of energy density and thermal pressure may affect significantly the cosmic expansion.

\end{abstract}
\maketitle

\section{\bf Introduction}
The cosmological observations developed in the last two decades indicate that the universe is undergoing a phase of cosmic acceleration started after the matter domination  \cite{riess,abaz,sper}.
It seems that some unknown energy components (dark energy) with negative pressure are responsible for this late-time acceleration \cite{cope}.
The simplest model which successfully explains the observational data is the $\Lambda$CDM ($\Lambda$-cold dark matter) model \cite{peri,jass,bah}. But in this model, the key question about the origin of the dark energy remains unanswered. Since the observed value of the cosmological constant (as the density of dark energy) is very small in comparison with the predicted vacuum energy of matter fields, it is not possible to attribute the dark energy directly to the quantum vacuum energy.
The origin and the nature of dark energy is still a mystery and its existence is beyond the domain of the standard model of particle physics and general relativity \cite{pad}.

Recently, an alternative approach to accommodate dark energy is modifying
the general theory of relativity on large scales. The motivation for modifying the gravitational part of the Einstein equation is not restricted to solve the cosmological problems. In fact, general relativity is not a renormalizable theory, and consequently to quantize the gravitational fields conventionally, the Einstein-Hilbert action needs to be supplemented by higher order curvature terms \cite{uti,stel}. Also, in string theory and when quantum corrections are taken into account, the effective gravitational action at low energy level admits higher order curvature invariants \cite{birr,buch,vilk}.

 Among these theories, scalar-tensor theories \cite{perr}, $f(R)$
gravity \cite{kle} , DGP braneworld gravity \cite{dvali} and string-inspired theories \cite{gross} are studied extensively.
Also, another theory in this context is scalar-Gauss-Bonnet gravity which is closely related
to the low-energy string effective action. In this proposal, the current acceleration of the
universe may be caused by mixture of scalar phantom and (or) potential/stringy effects \cite{noji}.
The coexistence of matter dominated and accelerating power law solutions for this theory has already been shown \cite{goheer}. It is also seen
that the Gauss-Bonnet gravity is less constrained than $f(R)$ gravity \cite{felic}.

On the other hand, as was first pointed out by Zeldovich \cite{zeld} and Hu \cite{hu}, the process
of matter creation in an expanding universe may phenomenologically be equivalent to effective negative bulk pressure. Therefore, in this context, the present accelerating stage may have two origins: the negative pressure resulting from the gravitational particle creation and the higher order terms of the gravitational sector.

 In these connections, the process
of matter creation in an expanding universe has been extensively discussed in the last five
decades {\cite{par}-\cite{ grib}.
The first thorough treatment of particle production by an external gravitational filed was given by Parker \cite{par, park}. However, the self-consistent macroscopic formulation of the matter creation process was put forward by Prigogine and coworkers \cite{prigo} and Calvao, Lima and Waga \cite{calv}.

In flat space-time, Poincare invariance is a guide which generally allows to identify a unique vacuum state
for the theory. However, in curved space-time, we do not have the Poincare symmetry.
The absence of Poincare symmetry in curved
space-time leads to the problem of the definition of particles and vacuum states. The problem
may be solved by using the method of the diagonalization of instantaneous Hamiltonian by
a Bogolyubov transformation, which leads to finite results for the number of created particles
\cite{grib}. In this direction, some works have been done in the context of modified gravity \cite{pereira, spereira, rsetare}.

In the present work, we investigate the particle production in a $f(G)$ theory for a flat and matter dominated
universe. It is proved that the total particle density always has a finite value. Therefore, the Bogolyubov transformations 
are well-defined and the Hilbert spaces spanned by the vacuum states at different times are unitarily equivalent \cite{muk}.
 The negative pressure resulting from the gravitational particle creation is also obtained for adiabatic processes, i.e. the processes in which the entropy per particle remains constant. In this case the entropy production density is entirely due to the increase of the number of particles \cite{prigo,calv,zimd,zimd2}. We show that, even if the higher order curvature terms of the Gauss-Bonnet gravity are ignored, this pressure can alone explain the present accelerating expansion. This result indicates that the pressure of the particle creation even in the presence of energy density and thermal pressure may affect significantly the cosmic expansion.
  Obviously, to reach a self-consistent model at least in a semiclassical framework one should take the back reaction effect of the particle creation into account, i.e. the gravitational equations and the particle creation equations must be solved simultaneously. But, since the coupling between the gravitational background and the density and pressure of particle creation is very complicated, it may be difficult. Then the result of the present paper might be viewed as the first approximation of the particle creation effect.

\section{Field equations}
We consider the following $f(G)$ action which describes Einstein's gravity coupled with
perfect fluid plus a function of the Gauss-Bonnet term \cite{nojir, nojiri}
\begin{equation}
S=\int d^{4}x \sqrt{-g}[\frac{1}{2 k^{2}}R+ f(G)+L_{m}],\label{eq1}
\end{equation}
where $k^{2}=8\pi G_{N}$ and the Gauss-Bonnet invariant is defined as follows
\begin{equation}
G= R^{2}- 4 R_{\mu\nu}R^{\mu\nu}+ R_{\mu\nu\lambda\sigma} R^{\mu\nu\lambda\sigma}. \label{eq2}
\end{equation}
By varying the action with respect to $g_{\mu\nu}$, it follows that
\begin{eqnarray}
0&=& \frac{1}{2k^{2}}(-R^{\mu\nu}+\frac{1}{2}g^{\mu\nu}R)+ T^{\mu\nu}+ \frac{1}{2}g^{\mu\nu} f(G)- 2f_{G}RR^{\mu\nu}+4f_{G}R^{\mu}_{\rho}R^{\nu\rho}\nonumber\\&-& 2 f_{G} R^{\mu\rho\sigma\tau}R^{\nu}_{\rho\sigma\tau}
-4f_{G}R^{\mu\rho\sigma\nu}R_{\rho\sigma}+2(\nabla^{\mu}\nabla^{\nu}f_{G})R-2g^{\mu\nu}(\nabla^{2}f_{G})R-4(\nabla_{\rho}\nabla^{\mu}f_{G})R^{\nu\rho}
\nonumber\\&-& 4(\nabla_{\rho}\nabla^{\nu}f_{G})R^{^{\mu\rho}}+4(\nabla^{2}f_{G})R^{\mu\nu}
+4g^{\mu\nu}(\nabla_{\rho}\nabla_{\sigma}f_{G})R^{\rho\sigma}-4(\nabla_{\rho}\nabla_{\sigma}f_{G})R^{\mu\rho\nu\sigma},\label{eq3}
\end{eqnarray}
where $f_{G}=f'(G)$ and $f_{GG}=f''(G)$. By using the metric of Friedmann-Robertson-Walker (FRW), we can obtain the first FRW equation
\begin{equation}
-\frac{3}{k^{2}}H^{2}+ G f_{G}- f(G)-24 \dot{G}H^{3}f_{GG}+\rho_{m}=0,_{} \label{eq4}
\end{equation}
where an over-dot denotes derivative with respect to time $t$ and Hubble
parameter $H$ is defined by $H=\frac{\dot{a}}{a}$. Also, using the
equation of state $P=w \rho_{m}$, the energy conservation law can be
expressed as
\begin{equation}
\dot{\rho}_{m}+ 3 H(1+w)\rho_{m}=0, \label{eq5}
\end{equation}
where $\rho_{m}$ is the matter density. Now, by assuming an exact power-law solution for the field equations as follows
\begin{equation}
a(t)= b t^{c}, \label{eq6}
\end{equation}
where $c$ and $b$ are positive real numbers, the Friedmann equation is exchanged as
\begin{equation}
\frac{4}{c-1}G^{2}f_{GG}+G
f_{G}-f_{G}-\frac{G^{\frac{1}{2}}}{k^{2}}(\frac{3c}{8(c-1)})^{\frac{1}{2}}+
\rho_{0}(\frac{G}{24c^{3}(c-1)})^{\frac{3}{4}c(1+w)}=0. \label{eq7}
\end{equation}
This is a differential equation for the function $f(G)$ in $G$ space. The general solution of this equation is obtained as
\begin{equation}
f(G)=-\frac{1}{2}\left[\sqrt{\frac{6c(c-1)}{k^{4}(c+1)^{2}}}G^{\frac{1}{2}}+A_{cw}G^{\frac{3}{4}c(1+w)}\right],
\label{eq8}
\end{equation}
where
\begin{equation}
A_{cw}=\frac{8\rho_{0}(c-1)(1382
c^{9}(c-1)^{3})^{-\frac{1}{4}c(1+w)}} {4+ c[3 c
(w+1)(w+\frac{4}{3})-15w-19]}. \label{eq9}
\end{equation}
As we see that a real valued solution for $f(G)$ requires the values
$c\leq 0$ or $c\geq 1$. In \cite{rast}, it is shown that only the
case $c>1$ leads to an accelerating universe. Also, in order to
avoid divergency in the Gauss-Bonnet term we have to keep $c$ and
$\omega$ away from the values for which $A_{cw}$ diverges according
to the following equation
\begin{equation}
4+ c [3 c(w+1)(w+\frac{4}{3})-15w-19]=0. \label{eq10}
\end{equation}
 For the case of matter dominated universe $(w=0)$, it must be supposed $c >
1$ and $c \neq\frac{19}{8}+\frac{\sqrt{345}}{8}$ ( by noting the
equation (\ref{eq10})). The Hubble parameter $
H=\frac{\dot{a}}{a}=\frac{c}{t}$ determines the actual age of
universe as $ t_{0}=\frac{c}{H_{0}}$ that is the order of $10^{9}$
years. Before studying the particle creation process, it is convenient to obtain the scale factor $a(t)$ in terms of the
conformal time $\eta$. The conformal time is defined as
\begin{equation}
\eta\equiv\int^{t}\frac{dt'}{a(t')}, \label{eq17}
\end{equation}
then we have
\begin{equation}
a(\eta)=\frac{B}{(-\eta)^{\frac{c}{c-1}}},\hspace{1cm}
B=[b (c-1)^{c}]^{\frac{1}{1-c}},\hspace{0.4cm} -\infty<\eta<0, \label{eq18}
\end{equation}
the early universe ( the past ) corresponds to $ \eta \rightarrow -\infty $ and the late universe ( the future ) corresponds to
$ \eta \rightarrow 0 $.
 To calculate the density of
particles per mode, we should determine the parameter $b$ in the numerator
of the scale factor $a(\eta)$. So, let us choose $b$ such that
$a(t_{0})\equiv a_{0}=1 $, for $t_{0}=\frac{c}{H_{0}}$, i.e., the
scale factor is normalized to unity for the present time. To satisfy
this condition we must have $ b=(\frac{H_{0}}{c})^{c}$. Using
(\ref{eq17}) we get the value of the present conformal time as
$\eta_{0}=-\frac{c}{c-1}\frac{1}{H_{0}}$. Therefore, the scale
factor becomes
\begin{equation}
a(\eta)= \frac{(-\eta_{0})^{\frac{c}{c-1}}}{(-\eta)^{\frac{c}{c-1}}}, \label{eq24}
\end{equation}

Although with the power-law solutions the evolution of the universe is basically restricted, these solutions help us to find some quantities analytically. But it is not the only motivation behind this choice. In fact, since these solutions are corresponding to the scaling solutions in $f(G)$ framework \cite{Uddin}, they play an important role in cosmology. They can be regarded as approximations to more realistic models and provide a framework for establishing the behavior of more general cosmological solutions \cite{Uddin}. Also, it has been proved that the scaling solutions are global attractors for a large class of cosmological models \cite{Copeland,Nun}. Therefore, the choice of such solutions is particularly relevant because in the Friedmann–Robertson–Walker backgrounds, they typically represent asymptotic or intermediate states in the full phase–space of the dynamical system representing all possible cosmological evolutions \cite{Goheer2}. They then enable us to determine the asymptotic behavior and stability of a particular cosmological background \cite{Liddle,Rubano,Copeland2,Tsujikawa,Steinhardt}.


In the following sections, we are going to study the particle creation process in this model.

\section{Scalar particle creation in $f(G)$ theory}
Generally, the field equation for the study of scalar particle creation in a spatially
flat Friedmann-Robertson-Walker geometry can be written as \cite{muk}
\begin{equation}\label{a1}
  {\cal X}''(\mathbf{x},\eta)-\nabla ^{2}{\cal X}(\mathbf{x},\eta)+(m^{2}a^{2}(\eta)-\frac{{a''(\eta)}}{a(\eta)}){\cal X}(\mathbf{x},\eta)=0,
\end{equation}
where the prime denotes derivative with respect to the conformal time $\eta$ and $\nabla ^{2}$ is the Laplacian.
Replacing the mode expansion
\begin{equation}\label{a2}
 {\cal X}(\mathbf{x},\eta)=\int\frac{d^3\mathbf{k}}{(2\pi)^{3/2}}\frac{1}{\sqrt{2}}[{a}_{k}^{-}{\cal X}^{*}_{k}(\eta)e^{i\mathbf{k}.\mathbf{x}}+
 {a}_{k}^{+}{\cal X}_{k}(\eta)e^{-i\mathbf{k}.\mathbf{x}}]
\end{equation}
in the field equation (\ref{a1}) implies the decoupled equations of motion for the modes ${\cal X}_{k}(\eta)$,
\begin{equation}
{\cal X}''_{k}+ \omega^{2}_{k}({\eta}){\cal X}_{k}(\eta)=0, \label{eq11}
\end{equation}
with
\begin{equation}
\omega^{2}_{k}({\eta})= k^{2}+ m^{2}_{eff} \hspace{1cm} and
\hspace{1cm} m^{2}_{eff}=m^{2}a^{2}(\eta)-\frac{{a''(\eta)}}{a(\eta)}.  \label{eq12}
\end{equation}

where ${\cal X}_{k}$ is the Fourier mode of the wave associated to the energy of the particle through the
frequency $\omega_{k}$, $m_{eff}$ represents an effective mass of the particle and the prime denotes derivative
with respect to the conformal time $\eta$.

Here, the quantization can be carried out by imposing equal-time commutation
relations for the scalar field ${\cal X}$ and its canonically conjugate momentum $\Pi\equiv{\cal X}'$,
namely $[{\cal X}(x,\eta), \Pi(y,\eta)]= i\delta(x-y)$, and by implementing secondary quantization in
the so-called Fock representation. After convenient Bogolyubov transformations, one obtains the transition amplitudes
for the vacuum state and the associated spectrum of the produced particles in a non-stationary background \cite{agrib,muk}. Usually,
the calculations of particle production deal with comparing the particle number at asymptotically early and late times,
or with respect to the vacuum states defined in two different frames and do not involve any loop calculation. Since
equation (\ref{eq11}) is a second order differential equation, we obtain two independent solutions.

To quantize the scalar field ${\cal X}(x,\eta)$ in the standard fashion by introducing the equal-time
commutation relations $[{\cal X}(\mathbf{x},\eta), {\cal X}'(\mathbf{y},\eta)]= i\delta(\mathbf{x}-\mathbf{y})$, each mode solution ${\cal X}_{k}$ must be normalized for all times according to
\begin{equation}
W_{k}(\eta)\equiv  {\cal X}^{'}_{k}(\eta){\cal X}^{*}_{k}(\eta)-{\cal X}_{k}(\eta){\cal X}^{*^{'}}_{k}(\eta)=2i. \label{eq13}
\end{equation}
If the vacuum state is defined as the lowest-energy eigenstate of the instantaneous Hamiltonian at time $\eta$, the mode decomposition ${\cal X}_{k}(\eta)$ corresponding to this vacuum state should satisfy the following conditions at time $\eta$ \cite{muk}
\begin{equation}\label{adeq1}
 {\cal X}_{k}(\eta)=\frac{e^{i\lambda}}{\sqrt{\omega_{k}(\eta)}},\hspace{1cm} {\cal X}^{'}_{k}(\eta)=i e^{i\lambda}\sqrt{\omega_{k}(\eta)},
\end{equation}
where $\lambda$ is an arbitrary real number. A mode function satisfying the above conditions defines a creation and annihilation set of operators $\hat{a}_{k}^{\pm}$ and then the vacuum state as the instantaneous lowest-energy state is the state annihilated by $\hat{a}_{k}^{-}$. In addition, the instantaneous Hamiltonian is diagonal in the eigenbasis of the occupation number operators $\hat{N}_{k}=\hat{a}_{k}^{+}\hat{a}_{k}^{-}$. But $\omega_{k}$ is not time-independent in a time-dependent gravitational background. Therefore, the mode function selected by the conditions (\ref{adeq1}) is time-dependent. It means that the vacuum states at different times differ from each other. However, these instantaneous vacuum states at different time are related by the Bogolyubov coefficients. If the mode function ${\cal X}_{k}(\eta)$ satisfies the conditions (\ref{adeq1}) at the initial time $\eta_{i}$ and if we suppose that the physical state is the instantaneous vacuum state corresponding to this mode function, then a straightforward calculation shows that
the final expression for the number density of created particles in the $k$ mode at time $\eta>\eta_{i}$ is \cite{agrib,muk}
\begin{equation}
N_{k}(\eta)= \frac{1}{4|\omega_{k}(\eta)|}|{\cal X}'_{k}(\eta)|^{2}+ \frac{|\omega_{k}(\eta)|}{4}|{\cal X}_{k}(\eta)|^{2}-\frac{1}{2}. \label{eq14}
\end{equation}
The proper density of particles per mode is given by
\begin{equation}
n_{k}(\eta)=\frac{N_{k}(\eta)}{a^{3}(\eta)}, \label{eq15}
\end{equation}
and the total number density of created particles is obtained by integrating overall the modes
\begin{equation}
n(\eta)=\int n_{k}(\eta) d^{3}\mathbf{k}. \label{eq16}
\end{equation}
It is worthwhile to note that the Bogolyubov transformation is well-defined only if the total particle density (\ref{eq16}) is finite. If this is not the case, the final vacuum state is not expressible as the normalized linear combination of the initial vacuum state and the excited states derived from it. In other words, the two Hilbert spaces spanned by these two vacuum states and their excited states are not unitarily equivalent.

\section{The creation of massless particles}
Here, by using equation (\ref{eq18}), we find $\frac{a''}{a}=[\frac{c (2c-1)}{(c-1)^{2}}] (-\eta )^{-2}$
that does not depend on the parameter $B$. The equation (\ref{eq11}) for mode function in the case of a massless particle
$(m=0)$ becomes
\begin{equation}
{\cal X}''_{k}(\eta)+ \left[k^{2}-\left(\frac{c (2c-1)}{(c-1)^{2}}\right)\frac{1}{\eta^{2}}\right]{\cal X}_{k}(\eta)=0. \label{eq19}
\end{equation}
According to the normalization condition (\ref{eq13}), the solution of this equation is given by
\begin{equation}
{\cal X}_{k}(\eta)=\sqrt{\frac{\pi |\eta |}{2}}[J_{\nu}(k|\eta|)+ i Y_{\nu}(k|\eta|)]=\sqrt{\frac{\pi |\eta |}{2}}H_{\nu}^{(1)}(k|\eta|),\hspace{1cm}
\nu\equiv\sqrt{\frac{1}{4}+\frac{c(2c-1)}{(c-1)^{2}}}, \label{eq20}
\end{equation}
where $J_{\nu}$ and $Y_{\nu}$ are respectively the Bessel functions of the first and second kind and $H_{\nu}$ is the Hankel function. The solution (\ref{eq20}) satisfies the lowest-energy conditions (\ref{adeq1}) at the initial time $(\eta \rightarrow -\infty)$ and has the correct asymptotic behaviour of the form
\begin{equation}
{\cal X}_{k} \rightarrow \frac{1}{\sqrt{k}} \exp (ik\eta+\delta), \label{eq22}
\end{equation}
where $\delta$ is a phase. This corresponds to plane waves for the
modes $k$ in the past. To calculate the spectrum of massless particles created during the
evolution of the universe by equation (\ref{eq14}), we have to firstly show that the total density of created particles (\ref{eq16}) has a finite value at all times. To show this one should prove that $N_{k}(\eta)$ tends to zero faster than $k^{-3}$ at $k\rightarrow\infty$. Employing the asymptotic expansion of the Hankel functions, i.e.
\begin{equation}\label{adeq2}
 H_{\nu}^{(1)}(k|\eta|)=\sqrt{\frac{2}{\pi k|\eta |}}\exp\{i[k|\eta|-(\nu+\frac{1}{2})\frac{\pi}{2}]\}(P_{\nu}(k|\eta|)+iQ_{\nu}(k|\eta|)),
\end{equation}
where
\begin{equation}\label{adeq3}
 P_{\nu}(z)+iQ_{\nu}(z)=\sum_{r=0}^{\infty}\frac{\Gamma(\nu+r+\frac{1}{2})}{r!\Gamma(\nu-r+\frac{1}{2})}(-2iz)^{-r},
\end{equation}
 it is not difficult to see that the first non-vanishing term of $N_{k}(\eta)$ at $k\rightarrow\infty$ is of order $k^{-4}$. Thus, $N_{k}(\eta)$ tends to zero faster than $k^{-3}$ at $k\rightarrow\infty$. Substituting solution (\ref{eq20}) into the relation (\ref{eq14}) and setting $c=10/3$ and $\eta=1$, we plot $k^3N_{k}(\eta)$ as a function of $k$ in Fig.1. It shows that $k^3N_{k}(\eta)$ tends to zero at $k\rightarrow\infty$.

\begin{figure}[h]
\centering
\includegraphics[scale=.6]{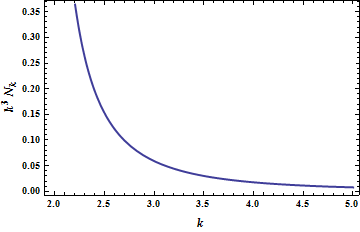}
\caption{This figure shows $k^3N_{k}(\eta)$ as a function of the mode $k$
with the parameters $c=\frac{10}{3}$ and $\eta=1$. It shows that $k^3N_{k}(\eta)$ tends to zero at $k\rightarrow\infty$. This asymptotic behaviour is necessary to prove that the total number density of created particles is finite.}
\end{figure}

 On the other hand, for each $k$ there is a critical
value of $\eta$ for which the number density of created particles grows abruptly
and diverges. Mathematically this
occurs because in the expression (\ref{eq14}) the frequency $\omega_{k}(\eta)$ appears in the
denominator, and when $k^{2}=\frac{c(2c-1)}{\eta^{2}(c-1)^{2}}$ we have $\omega_{k}(\eta)\rightarrow 0$. Physically, the
significance of this divergence is that for values of $ k^{2} < \frac{c(2c-1)}{\eta^{2}(c-1)^{2}}$, the frequency $\omega_{k}^{2}$ becomes negative, consequently the state of minimum energy
and the quantum vacuum are not well-defined in these cases, and the creation ceases for these values. Therefore, the total number density of particle should be determined as
\begin{equation}\label{adeq4}
  n(\eta)=\frac{1}{a^{3}(\eta)}\int_{k^{2}=\frac{c(2c-1)}{\eta^{2}(c-1)^{2}}}^{\infty} N_{k}(\eta) d^{3}\mathbf{k}.
\end{equation}
Although the first term of $N_{k}(\eta)$, i.e. $\frac{|{\cal X}'_{k}(\eta)|^{2}}{4|\omega_{k}(\eta)|}=\frac{|{\cal X}'_{k}(\eta)|^{2}}{4\sqrt{k^2-\frac{c(2c-1)}{\eta^{2}(c-1)^{2}}}}$, tends to infinity at $k^2\rightarrow\frac{c(2c-1)}{\eta^{2}(c-1)^{2}}$, its integral is finite because the following integral is convergent:
\begin{equation}\label{adeq5}
  \int_{a}^{b>a}\frac{dx}{\sqrt{x^2-a^2}}=\ln(\frac{b+\sqrt{b^2-a^2}}{a}).
\end{equation}
Then the total number density of particles has a finite value at all times. In Fig.2, the dimensionless total number density of particles defined as $\eta_{0}^3 n(\eta)$ is displayed versus $\frac{\eta}{|\eta_{0}|}$.

\begin{figure}[h]
\centering
\includegraphics[scale=.6]{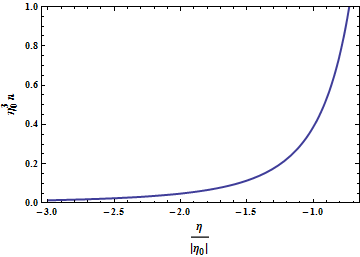}
\caption{ The figure presents the dependence of the dimensionless total number density of particles $\eta_{0}^3 n(\eta)$ on $\frac{\eta}{|\eta_{0}|}$.}
\end{figure}

Now, using (\ref{eq14}) and (\ref{eq20}), we can calculate the evolution
of particle density $N_{k}$ for each mode $k$. Fig.3 shows the
density of massless particle created during the evolution of the
universe as a function of $\frac{\eta}{|\eta_{0}|}$ for different
values of mode $k$. The value $\frac{\eta}{|\eta_{0}|}=-1$ represents
the present time. In the past $(\eta\rightarrow -\infty)$ the
number density is zero for all modes, but it grows throughout evolution.

\begin{figure}[h]
\centering
\includegraphics[scale=.6]{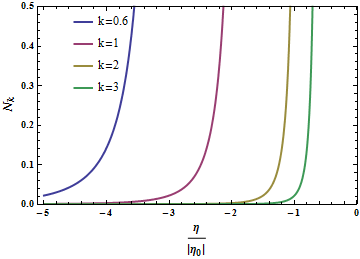}
\caption{This figure shows the evolution of density of created particle $N_{k}$ for four different values of the mode $k$
with the parameter $c=\frac{10}{3}$ (for comparing with results of \cite{spereira}, we take $c=\frac{10}{3}$).}
\end{figure}

By using this figure, as mentioned earlier,  we can see that for each $k$ there is a critical
value of the conformal time $\eta$ for which the number density of created particles grows abruptly
and diverges. For $k = 1$ this occurs in the
past when $\frac{\eta}{|\eta_{0}|}\approx -2.05 $, for $k = 2$ we have $\frac{\eta}{|\eta_{0}|}\approx -1.02 $, very close to the
present time. But for $k = 3$ this will only occur in the future, $\frac{\eta}{|\eta_{0}|}\approx -0.7$.
 A value of $k < 1$ is also shown in this figure.

\section{Pressure of particle creation}
It is not difficult to show that the interactions wherein particle number conservation violated, including the gravitational particle creation, may lead to an effective negative pressure \cite{prigo}.
From the first law of thermodynamics and Euler's relation for an open system in which the particle number $N$ is time dependent, it follows that
\begin{equation}\label{pc1}
  d(\rho V)+PdV-\frac{\rho+P}{n}d(nV)-TNd(\frac{S}{N})=0,
\end{equation}
where $\rho=E/V$ and $n=N/V$. The above relation is known as the Gibbs relation. The Gibbs relation can be also written as
\begin{equation}\label{pc2}
  d\rho-\frac{\rho+P}{n}dn=nT d(\frac{S}{N}).
\end{equation}
If the entropy per particle $S/N$ is constant, i.e. the entropy production is entirely due to the increase of the number of particles, the Gibbs relation implies that
\begin{equation}\label{pc3}
  d(\rho V)+PdV-\frac{\rho+P}{n}d(nV)=0,
\end{equation}
or equivalently
\begin{equation}\label{pc4}
  \dot{\rho}=(\rho+P)\frac{\dot{n}}{n},
\end{equation}
where an over-dot denotes the derivative with respect to time. Now, by defining a supplementary pressure $P_{c}$
\begin{equation}\label{pc5}
  P_{c}=-\frac{\rho+P}{n}\frac{d(nV)}{dV}.
\end{equation}
equation (\ref{pc3}) can be rewritten as
\begin{equation}\label{pc6}
  d(\rho V)=-(P+P_{c})dV,
\end{equation}
It means that the creation of matter corresponds to a supplementary pressure $P_{c}$ which must be considered as a part of the total pressure $P_{t}$ entering into the matter part of the gravitational equations \cite{prigo,calv,zimd,zimd2}, i.e.
\begin{equation}\label{pc7}
  P_{t}=P+P_{c}.
\end{equation}
In the case of an isotropic and homogeneous universe, one can set $V=a^{3}(t)$, then
\begin{equation}\label{pc8}
  P_{c}=-(\rho+P)(\frac{\dot{n}}{3Hn}+1).
\end{equation}
Before determining the pressure of particle creation, it should be noted that the distribution $\omega_{k}(\eta)n_{k}(\eta)$ as a function of $k$ does not correspond to a thermodynamic equilibrium state. Thus, similar to the Gamow condition, we should assume that the transition rate to an equilibrium state is faster than the particle production rate and the expansion rate of the universe. Using equations (\ref{eq24}) and (\ref{pc8}), in Fig.4 the behavior of $\frac{P_{c}}{\rho+P}$ as a function of $\frac{\eta}{|\eta_{0}|}$ is displayed.

\begin{figure}[h]
\centering
\includegraphics[scale=.6]{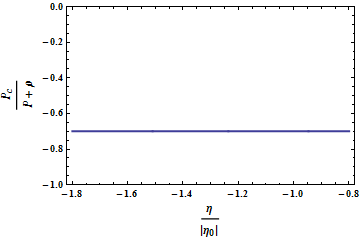}

\caption{This figure shows the evolution of $P_{c}/(\rho+P)$ as a function of $\frac{\eta}{|\eta_{0}|}$. It is proved that this function has a constant value.}
\end{figure}
It shows that the pressure of particle creation has a negative value as it is expected. It also seems that $\frac{P_{c}}{\rho+P}$ has a constant value. This constancy is not an amazing result because from equations (\ref{eq14}), (\ref{eq19}), (\ref{eq20}) and (\ref{adeq4}) it follows that $n(\frac{\eta}{\alpha})=\alpha^{3}n(\eta)$ for all $\eta<0$ and $\alpha>0$. Then it is not difficult to prove that $n(\eta) \propto \eta ^{-3}$, which implies $\frac{\dot{n}}{3Hn}=const.$

To compare the effect of this negative pressure on the acceleration of expansion with the effect of energy density and thermal pressure which always have positive values, we can for simplicity neglect the contribution of the higher order curvature terms of the Gauss-Bonnet gravity. So it is enough to restrict ourselves to general relativity.
In this case, the Einstein equations yield \cite{lima}
\begin{equation}\label{pc9}
 8\pi G_{N}\rho=3\frac{\dot{a}^2}{a^2}+3\frac{k}{a^2},
\end{equation}
and
\begin{equation}\label{pc10}
8\pi G_{N}(P+P_{c})= -2\frac{\ddot{a}}{a}-\frac{\dot{a}^2}{a^2}-\frac{k}{a^2},
\end{equation}
which imply
\begin{equation}\label{pc11}
  2\frac{\ddot{a}}{a}=-8\pi G_{N}(P_{c}+P+\frac{\rho}{3}).
\end{equation}
Taking $P=\rho/3$, from equation (\ref{pc8}) and Fig.4, it follows that the right hand side of above equation has a positive value and therefore, $\ddot{a}>0$.
 This result indicates that the pressure of the particle creation even in the presence of energy density and thermal pressure may affect significantly the cosmic expansion.

 Obviously, to reach a self-consistent model at least in a semiclassical framework one should take the back reaction effect of the particle creation into account, i.e. the gravitational equation (\ref{eq4}) and the particle creation equation (\ref{eq19}) must be solved simultaneously. But, since the coupling between the gravitational background and the density and pressure of particle creation is very complicated, it may be difficult. Then the result of the present paper might be viewed as the first approximation of the particle creation effect.

\section{Conclusions}
We have investigated the problem of massless particle creation in a
$f(G)$ theory for a matter dominated universe. We have assumed an
exact power-law solution for the scale factor of universe, which
leads us to an accelerated expanding universe.
 The amount of particles created with $k < 1$ is steadily
increasing in the past, although the creation of such modes stop
abruptly in the past (Fig 3.). This shows that in the past a huge amount of
particles with low $k$ were created. These results perfectly agree
with one of that obtained in studying quantum effect in the context of
$f(R)$ gravity \cite{spereira}. 
It has been also proved that the total particle density always has a finite value. Therefore, the Bogolyubov transformations are well-defined and the Hilbert spaces spanned by the vacuum states at different times are unitarily equivalent. In addition, we have shown that the pressure of particle creation has negative value as it is expected. This pressure even in the presence of energy density and thermal pressure may affect significantly the cosmic expansion.



\begin{thebibliography}{99}
\bibitem{riess} A. G. Riess et al. [Supernova Search Team Collaboration], Astron. J. 116, 1009 (1998)
[astro-ph/9805201];\\
S. Perlmutter et al. [Supernova Cosmology Project Collaboration], Astrophys. J. 517, 565 (1999)
[astro-ph/9812133];\\
P. Astier et al., Astron. Astrophys. 447, 31 (2006) [astro-ph/0510447].
\bibitem{abaz} K. Abazajian et al. [SDSS Collaboration], Astron. J. 128, 502 (2004) [astro-ph/0403325];\\
K. Abazajian et al. [SDSS Collaboration], Astron. J. 129, 1755 (2005) [astro-ph/0410239].
\bibitem{sper} D. N. Spergel et al. [WMAP Collaboration], Astrophys. J. Suppl. 148, 175 (2003) [astro-ph/0302209];\\
D. N. Spergel et al., astro-ph/0603449.
\bibitem{cope} E. J. Copeland, M. Sami and Shinji Tsujikawa, Int. J. Mod. Phys. D 15 (2006) 1753; Y-F Cai, E. N.
Saridakis, M. R. Setare, J-Q. Xia, Phys. Rep. 493, 1, (2010).
\bibitem{peri} L. Perivolaropoulos,
arXiv:astro-ph/0601014.
\bibitem{jass} H. Jassal, J. Bagla and T. Padmanabhan,
    Phys.Rev.D 72(2005)103503 [arXiv:astro-ph/0506748].
\bibitem{bah} N. A. Bahcall, J.P. Ostriker, S. Perlmutter, P.J. Steinhardt,
Science 284(1999)1481.
\bibitem{pad} T. Padmanabhan, Phys. Rept. 380, 235 (2003); P. J. E. Peebles and B. Ratra, Rev. Mod. Phys. 75, 559 (2003); J. A. S. Lima, Braz. J. Phys. 34, 194 (2004); V. Sahni and A. Starobinsky, Int. J. Mod. Phys. D 15, 2105 (2006).
\bibitem{uti} R. Utiyama and B. S. DeWitt,
J. Math. Phys. 3(1962)608.
\bibitem{stel} K. S. Stelle, 
 Phys. Rev. D16(1977)953.
\bibitem{birr} N. D. Birrell and P. C. W. Davies, Quantum Fields in Curved Space (Cambridge University
Press, Cambridge, 1982); S. A. Fulling, Aspects of Quantum Field Theory in Curved Spacetime
(Cambridge University Press, Cambridge, 1989).
\bibitem{buch} I. L. Buchbinder, S. D. Odintsov, and I. L. Shapiro, Effective Actions in Quantum Gravity, IOP Publishing, Bristol, 1992.
\bibitem{vilk} G. A. Vilkovisky,
Class. Quant. Grav. 9(1992)895.
\bibitem{perr} F. Perrotta, C. Baccigalupi and S. Matarrese, Phys. Rev. D. B61 (2000) 023507; B. Boisseau, G.
Esposito-Farese, D. Polarski and A. A. Starobinsky, Phys. Rev. Lett. 85 (2000) 2236.
\bibitem{kle} H. Kleinert, H .J Schmidt, Gen. Rel. Grav.34, 1295, (2002); S. Capozziello, V. F. Cardone, S. Carloni
and A. Troisi, Int. J. Mod. Phys. D 12 (2003) 1969-1982, [arXiv:astro-ph/0307018]; S. Nojiri and S.
D. Odintsov, AIP Conf. Proc. 1115 (2009) 212-217, [arXiv:0810.1557]; T. P. Sotiriou and V. Faraoni,
[arXiv:/0805.1726]; M. R. Setare, Int. J. Mod. Phys. D 17 (2008) 2219, [arXiv:0901.3252].
\bibitem{dvali} G. Dvali, G. Gabadadze and M. Porrati, Phys. Lett. B 485 (2000) 208, [hep-th/0005016].
\bibitem{gross} D. J. Gross and J. H. Sloan, Nucl. Phys. B 291 (1987) 41; C. Charmousis and J. F. Dufaux, Class.
Quant. Grav. 19 (2002) 4671, [arXiv:hepth/ 0202107]; S. C. Davis, Phys. Rev. D 67 (2003) 024030,
[arXiv:hep-th/0208205].
\bibitem{noji} S. Nojiri, S. D. Odintsov and M. Sasaki, Phys. Rev. D 71, 123509 (2005); S. Nojiri, S. D.
Odintsov and M. Sami, Phys. Rev. D 74, 046004, (2006); B. M. N. Carter, I. P. Neupane,
Phys. Lett. B638, 94, (2006); B. M. N. Carter, and I. P. Neupane, JCAP 0606, 004, (2006);
J. W. Moffat, and V. T. Toth . arXiv:0710.0364 [astro-ph].
\bibitem{goheer} N. Goheer, R. Goswami, P. Dunsby, and K. Ananda, Phys. Rev. D79, 121301(R) (2009).
\bibitem{felic} A. De Felice and S. Tsujikawa, Phys. Lett. B 675, 1 (2009).
\bibitem{zeld} Ya. B. Zel'dovich, Pis'ma Zh. Eksp. Teor. Fiz. 12, 443 (1970)
[JETP Lett. 12, 307 (1970)].
\bibitem{hu} B. L. Hu, Phys. Lett. 90A, 375 (1982).
\bibitem{par} L. Parker, Phys. Rev. Lett. 21, 562 (1968).
\bibitem{park} L. Parker, Phys. Rev. 183, 1057 (1969); Phys. Rev. D 3, 346 (1971); Phys. Rev. Lett. 28,
705 (1972); Phys. Rev. D 7, 976 (1973).
\bibitem{prigo} I. Prigogine et al., Gen. Rel. Grav., 21, 767 (1989).
\bibitem{calv} M. O. Calvao, J. A. S. Lima, I. Waga, Phys. Lett. A162, 223 (1992).

\bibitem{agrib} A. A. Grib, S. G. Mamayev and V. M. Mostepanenko, Vaccum Quantum effects in Strong Fields (Friedmann Laboratory Publishing, St. Petesburg, 1994).
\bibitem{muk} V. F. Mukhanov and S. Winitzki, Introduction to Quantum Effects in Gravity, Cambridge
University Press, Cambridge, 2007.
\bibitem{zimd} W. Zimdahl and D. Pavon, Phys. Lett. A 176, 57 (1993).
\bibitem{zimd2} W. Zimdhal, D. J. Schwarz, A. B. Balakin and D. Pavon,
Phys. Rev. D 64, 063501 (2001).
\bibitem{lima} J. A. S. Lima, F. E. Silva and R.C. Santos, Class. Quantum
Grav. 25(2008)205006.
\bibitem{zel} Ya. B Zel'dovich, Pisma Zh. Eksp. Teor. Fiz. 12, 443 (1970), (English transl. JETP 12, 307
(1970)); A. A. Grib, S. G. Mamayev and V. M. Mostepanenko, Gen. Rel. Grav., 7, 535 (1975);
A. A. Grib and Yu. V. Pavlov, [gr-qc/0505140]; Grav. Cosmol. 11, 119 (2005); Grav. Cosmol.
12, 159 (2006).
\bibitem{gri} L. P. Grishchuk, Class. Quant. Grav. 10, 2449 (1993); M. R. G. Maia, Phys. Rev. D 48, 647
(1993); M. R. G. Maia and J. D. Barrow, Phys. Rev. D 50, 6262 (1994); M. R. G. Maia and
J. A. S. Lima, Phys. Rev. D 54, 6111 (1996).
\bibitem{grib} A. A. Grib and S. G. Mamayev, Yad. Fiz. 10, 1276 (1969)[English transl.: Sov. J. Nucl. Phys.
10, 722 (1970)]. See also Yu. V. Pavlov, Theor. Math. Phys. 126, 92 (2001), [gr-qc/0012082].
\bibitem{pereira} S. H. Pereira, C. H. G. Bessa and J. A. S. Lima, Phys. Lett. B690: 103-107 (2010), arXiv:0911.0622 [astro-ph].
\bibitem{spereira} S. H. Pereira, J. C. Z. Aguilar and E. C. Romão, arXiv:1108.3346.
\bibitem{rsetare} M. R. Setare and M. J. S. Houndjo, arXiv:1111.2821 [physics.gen-ph].
\bibitem{nojir} S. Nojiri, S. D. Odintsov, Phys. Lett. B631, 1 (2005).
\bibitem{nojiri} S. Nojiri, S. D. Odintsov, O. G. Gorbunova, J. Phys. A39, 6627 (2006).
\bibitem{rast} A. R. Rastkar, M. R. Setare and F. Darabi, Astrophys Space Sci (2012) 337:487-491, arXiv:1104.1904.
\bibitem{Uddin} K. Uddin, J. E. Lidsey and R. Tavakol, Gen. Rel. Grav 41, 2725-2736 (2009),	arXiv:0903.0270 [gr-qc].
\bibitem{Copeland} E. J. Copeland, A. R. Liddle and D. Wands, Phys. Rev. D 57, 4686 (1998), arXiv:gr-qc/9711068.
\bibitem{Nun} A. Nunes and  J. P. Mimoso, Phys.Lett. B488 (2000) 423-427, arXiv:gr-qc/0008003.
\bibitem{Goheer2} N. Goheer, J. Larena and P. K. S. Dunsby, Phys. Rev. D80:061301 (2009), arXiv:0906.3860 [gr-qc].
\bibitem{Liddle} A. R. Liddle and R. J. Scherrer, Phys. Rev. D 59, 023509 (1999), arXiv:astro-ph/9809272.
\bibitem{Rubano} C. Rubano and J. D. Barrow, Phys. Rev. D 64, 127301 (2001), arXiv:gr-qc/0105037.
\bibitem{Copeland2} E. J. Copeland, S. J. Lee, J. E. Lidsey and S. Mizuno, Phys. Rev. D 71, 023526 (2005), arXiv:astro-ph/0410110.
\bibitem{Tsujikawa} S. Tsujikawa and M. Sami, Phys. Lett. B 603, 113 (2004), arXiv:hep-th/0409212.
\bibitem{Steinhardt} P. J. Steinhardt, L. M. Wang and I. Zlatev, Phys. Rev. D 59, 123504 (1999), arXiv:astro-ph/9812313.

\end{thebibliography}
\end{document}